# Context Aware Adaptable Applications - A global approach

**Marc DALMAU, Philippe ROOSE, Sophie LAPLACE**

**LIUPPA, IUT de Bayonne**
**2, Allée du Parc Montaury**
**64600 Anglet**
**FRANCE**
Mail : *{FirtsName.LastName@iutbayonne.univ-pau.fr}*

## Abstract

Actual applications (*mostly component based*) requirements cannot be expressed without a ubiquitous and mobile part for end-users as well as for M2M applications (*Machine to Machine*). Such an evolution implies context management in order to evaluate the consequences of the mobility and corresponding mechanisms to adapt or to be adapted to the new environment. Applications are then qualified as context aware applications.

This first part of this paper presents an overview of context and its management by application adaptation. This part starts by a definition and proposes a model for the context. It also presents various techniques to adapt applications to the context: from self-adaptation to supervised approached.

The second part is an overview of architectures for adaptable applications. It focuses on platforms based solutions and shows information flows between application, platform and context. Finally it makes a synthesis proposition with a platform for adaptable context-aware applications called *Kalimucho*. Then we present implementations tools for software components and a dataflow models in order to implement the *Kalimucho* platform.

## 1. Introduction

Actual applications (*mostly component based*) requirements cannot be expressed without a ubiquity and mobile part for end-users as well as for M2M applications (*Machine to Machine*). Such an evolution implies context management in order to evaluate the consequences of the mobility and corresponding mechanisms to adapt or to be adapted to the new environment. Mobile computing and next, ubiquitous computing, focuses on the study of systems able to accept dynamic changes of hosts and environment [33] . Such systems are able to adapt themselves or to be adapted according to their mobility into a physical environment. That implies dynamic interconnections, and the knowledge of the overall context. Due to the underlying constraints (mobility, heterogeneity, etc.), the management of such applications is complex and requires considering constraints as soon as possible and having a global vision of the application.

Adaptation decision can be fully centralized (*A - Figure 1*) or fully distributed with all intermediary positions (*B&C - Figure 1*). The consequence is the level of autonomy of decision as well as the level of predictability. Obviously, the autonomy increases with decentralized supervision. Reciprocally, the complexity increases with the autonomy (problems of predictability, concurrency, etc.).

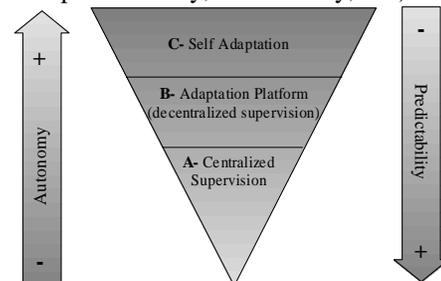

Figure 1 : Means of adaptation

Self-adaptable applications need to access to context information. This access can be active if the application captures itself the context (see A - Figure 1), or passive if an external mechanisms gives it access to the context (see B - Figure 1).

Nevertheless, with mobile peripherals and the underlying connectivity problems, a fully centralized supervision is not possible. A pervasive supervision [29] appears is a good solution and allows managing complexity, predictability while keeping the advantages of autonomy.

In order to be context-aware, applications need to get information corresponding to three adaptation types: data, service and presentation. The first one deals with "raw data" and its adaptations to provide complete and formatted information. Service adaptation deals with the architecture of the application and with dynamic adaptation (*connection/disconnections/migration of* *components composing the application*). It allows adapting the application in order to respect the required QoS. Presentation deals with HCI (*not addressed in this paper*).

Here is a global schema of an adaptable application:

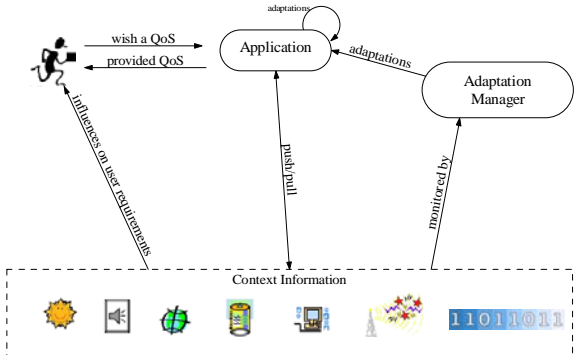

Figure 2 : Adaptable applications

Whereas [34] [35] do not make distinction between context oriented and application oriented data (functional data), we think that such a distinction makes design easier and offers a good separation of concerns [36] .

## 2.  What is context?

### 2.1  Definition and model

The origin of the term « context awareness » if attributed to Schilit and Theimer [42] . They explain that it is « *the capacity for a mobile application and/or a user to discover and react to situations modifications* ». Many other definitions can be found in [43] . The application context is the situation of the application so the context is a set of external data that may influence the application [36] .

A context management system can interpret the context and formalize it in order to make a high level representation. It creates an abstraction for the entities reacting to situations evolutions, they can be applications [35] , platforms, middlewares, etc. In order to make such abstractions, a three layered taxonomy can be organized as shown in Figure 3:

The first layer deals with context information capture. The first type of context is called Environmental: this is the physical context. It represents the external environment where information is captured by sensors. This information is about light, speed, temperature, etc. The second type, called User, gives a representation of users interacting with the application (language, localization, activity, etc.). This is the user profile. The third one deals with **Hardware**. Most probably, the more "classical" one; it gives information on available resources (memory, % CPU, connections, bandwidth, debit, etc.).  It also gives information as displays resolutions, type of the host (PDA,

Smartphone, Mobile Phone, PC, etc.). The third one is the **Temporal** context. It preserves and manages the history (date, hour, synchronizations, actions performed, etc.). The last one is called **Geographic** and gives geographical information (GPS Data, horizontal/vertical moving, speed, etc.).

The second layer, called « context management » [44] [45] is based on the previous layer representations. It provides **services** specifying the software environment of the application (platform, middleware, operating system, etc.). The **Storage** of context data in order to allow services retrieving them, the **Quality** giving a measure about the service itself or data processed and the **Reflexivity** allowing to represent the application itself. The **localization** manages geographic information in order to locate entities, predict their displacements.

The last layer proposes mechanisms to permit the adaptation to the context. We will find several mechanisms in order to react to contextual events. The first one is the software component **Composition,** the second one is the **Migration** in order to move entities and the last one, the **Adaptation** to ensure the evolution of the application. This last point is no-functional, the middleware manages it, it can depend on a user profile or on rules provided by the user. The **polymorphism** facilitates the migration of entities and their adaptation to various hosts (with more or less constraints).

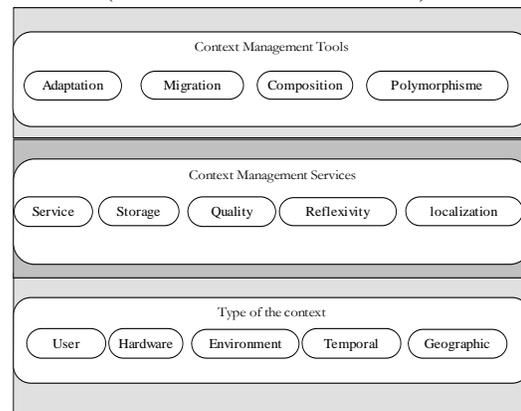

Figure 3 : Taxonomy of context

We propose a context model able to design any context information. This model (called *Context Object*) provides information needed by entities managing the application. Some information defines the context (its nature) whereas others define its validity. The nature of the context can be [34] :

- *User* (people) as his preferences,
- *Hardware* (things)  as network,
- *Environment* (places) as temperature, sunlight, sound, movement, air pressure, etc. It is the physical context. It represents  the  external  environment  from  where information is captured by sensors. It deals with users' environment [36] as well as hardware environment.

Such information is called *ContextInformation* and we call *InformationValidity* the validity area of a *ContextInformation (example:* old information or information which source is very far can be useless). *InformationValidity* is:

- *Temporal*: Temporal information can be a date or time used as timestamp. Time is one aspect of durability so it is important to date information as soon as it is produced. This temporal information allows making historical report of the application and defining the validity (*freshness*) of *ContextInformation* [40] . This freshness is the time since the last sensor reads it. Ranganathan defines a temporal degradation function that degrades the confidence of the information.
- *Spatial*: it is the current location (the host (identity) or the geographic coordinates (GPS)) and makes possible to distinguish local and remote context
- *Confidence information*: how precise is the sensor
- *Information ownership*: in some application hosted on a SmartPhone for example, privacy is very important, therefore, each information has to be stamped with its owner (*the producer*).

Let's notice that some information is strongly coupled as freshness and confidence whereas others are defined using application data as ownership. That is the reason why [39] identified physical, virtual (data source from software application or services) and logical sensors (combine physical and virtual sensors with additional information)

Depending on the application, one information type could be a *ContextInformation* or a *ValidityInformation*. For example, location can be a *ContextInformation* for a user in a forest or can be a *ValidityInformation* for the sensor network that supervises temperature and air pressure measurement.

According to this model, we organize all the characteristics of context information that define type, value, time stamp, source, confidence and ownership [37] or user, hardware, environment and temporal [45] [46] **Error! Reference source not found.**. In order to structure such contextual information, we proposed a meta-model structuring *ContextInformation* and *ValidityInformation* (see Figure 4).

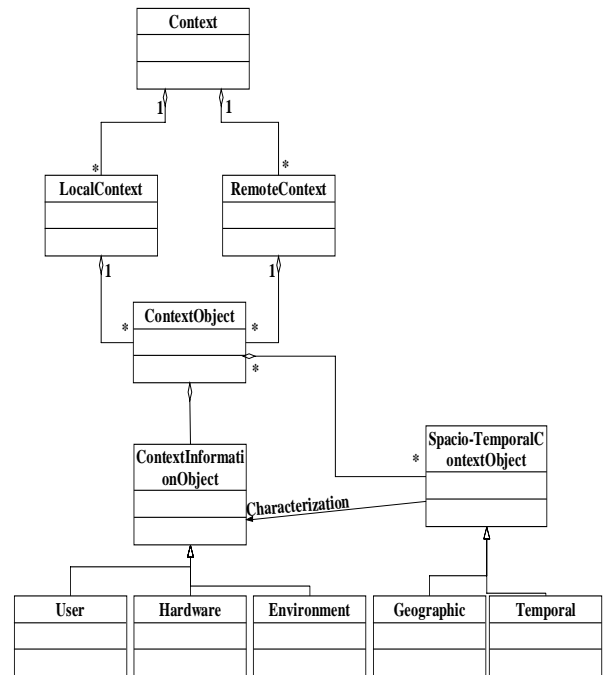

Figure 4 : Context class diagram

## 2.2 Context and applications

Since several years, the natural evolution of applications to distribution shows the need of more than only processing information. Traditionally, applications are based on input/output, i.e. input data given to an application produces output data. This too restrictive approach is now old fashioned [48] . Data are not clearly identified, processes does not only depend on provided data but depend also on data such the hour, the localization, preferences of the user, the history of interactions, etc. in a word the context of the application. We can find a representative informal definition in [49] "The execution context of an application groups all entities and external situations that influence on the quality of service/performances (qualitative & quantitative) as the user perceives them".

Designers and developers had to integrate the execution environment into their applications. This evolution allows applications to be aware of the context, then to be context-sensible and then to adapt their processes and next to dynamically reconfigure themselves in order to react as well as possible to demands. This is evidence, but to adapt itself to the context, the application needs to have a good knowledge of it and of its evolutions.

With a research point of view, context needs a vertical approach. All research domains/layers manage contextual information. Many works deal with its design, management, evaluation, etc. Its impact is wide: Re-engineering, HCI, Grid, Distributed Applications,

Ubiquitous Computing, Security, etc. But to be honest, the context it not a new concept in computer science! Since the early 90's, Olivetti Research Center with the ActiveBadge [Harter, 1994] and most of all, with a lot a regrets, the Xerox PARC with the PARCTab System [51] gave the bases of modern context aware applications.

In order to be aware of the context, the following architecture (see Figure 5) is "classical". An example can be found in [46] . It can be summarized as a superposition of layers. Each of them matches to a contextual information acquisition process, a contextual information management and an adaptation of the application to the context (as defined in Figure 3).

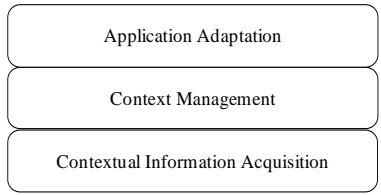

Figure 5 : Architectural layers of context aware applications

According to Figure 5, context management do imply to have dynamic applications in order to adapt them to variations of the context and so to provide a quality of service corresponding to current capabilities of the environment (application + runtime).

## 3. Context aware applications

Context aware applications are tightly coupled to mobiles devices and ubiquitous computing in the meaning of "machines that fit the human environment, instead of forcing humans to enter theirs" [1] . These applications are able to be aware of their execution context (user, hardware and environment) and its evolutions. They extract information from the context about geographical localization, time, hardware conditions (network, memory, battery, etc.) as well as about users.

Interactions between an application and its context can then be represented by two information flows (Figure 6):

- Application captures information from its context
- Application acts its context

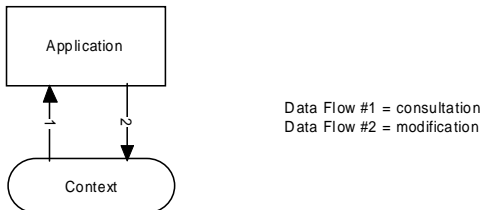

Data Flow #1 = consultation
Data Flow #2 = modification

Figure 6 : Context aware application

The means operated to realize both data flows of the Figure 6 depend on types of context (Table 1). They are system and network primitives for hardware context (resource allocation, connections, consultation of available resources, etc.). The user's context is captured through the interfaces and the information system (user's profile description files). At last, environmental context can be captured through sensors and modified by actuators.

| Flow | Type of context | | |
|------|-----------------|---|---|
| | Hardware | User | Environment |
| #1 | System and network primitives | Interfaces and information system | Sensors |
| #2 | Resource allocation | Interfaces | Actuators |

Table 1 : Means of interaction between application and context

However, even if it is possible to design limited applications according to the use of contextual information, the main interest is to be able to adapt the behavior of the applications to the context evolutions. Particularly, the increasing use of mobile and limited devices implies the deployment of adaptable applications. Such approach allows having a quality of service management (functional and non-functional services as energy saving for example).

### 3.1 Adaptable context aware applications

Adding adaptation to context aware applications means the addition of a new interaction corresponding to the influence that the context has on the application. That is the property for the application to adapt itself to the context (Figure 7).

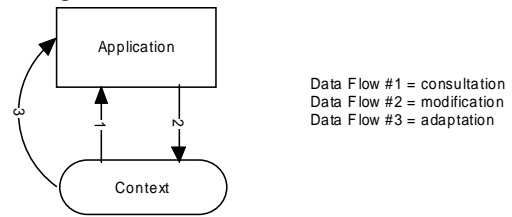

Data Flow #1 = consultation
Data Flow #2 = modification
Data Flow #3 = adaptation

Figure 7 : Adaptable Context Aware Application

Achievement of a context aware application can be done:

- By self adaptation
- By supervision

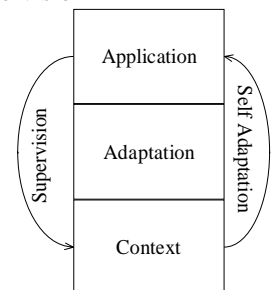

Figure 8 : Supervision vs Self Adaptation: a global vue.

### 3.1.1   Self adaptation

Such systems are expected to dynamically self-adapt to accommodate resource variability, change of user needs and system faults. In [27] , self-adaptive applications are described as useful for pervasive applications and sensor systems. Self-adaptive applications mean that adaptations are managed by the applications itself. It evaluates its behavior, configuration, and with distributed application, its deployment. The application captures the context (flow #1) and therefore adapts its behavior (data flow #3). The activity of the application modifies the context (flux #2). This approach, represented in Figure 7, raises the essential problem of accessing to distant context information. Indeed, through the interactions described in Table 1 it is only possible for the application to interact with its local context. In order to get or modify distant contextual information, the designer of the application has to set up specific services on the different sites of its application. It becomes necessary to set up many non functional mechanisms that strongly increases the complexity of the application and are difficult to maintain up to date.

Moreover self-adaptive solutions imply to have a planning and an evaluation part at runtime and a control loop. In order to make the evaluation, such application needs components description, as well as software description, structure and various alternatives, i.e. various assembling configurations.

Such solutions do not simplify the separations of concern, and so increase the practical viability of the application and its maintainability and possible evolutions. Moreover, with ubiquitous and heterogeneous environments, such generic solutions are not suitable to exploit the potential of hosts [28] . That is the reasons why most systems tend to solve these problems using platforms.

### 3.1.2   Supervised adaptation

In these approaches a runtime platform interfaces the application and the context.  It allows then access to distant context. The application only senses the context (flow #1) by means of the middleware of the platform. The application can modify the context and the platform itself (flow #2). Both the application and the platform adapt themselves to the context (flow #3). This kind of organization is shown in Figure 9.

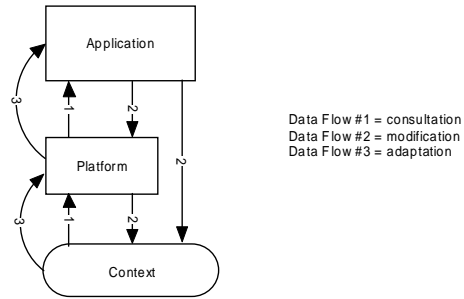

Data Flow #1 = consultation
Data Flow #2 = modification
Data Flow #3 = adaptation

Figure 9 : Adaptable Context Aware Application with platform

Recent works as Rainbow use a closed-loop control based on external models and mechanisms to monitor and adapt system behavior at run time to achieve various goals [32] , such solution is closed to the use of pervasive supervision. In order to implement such a solution, we need a distributed platform on all heterogeneous hosts. Such architecture allows to capture local context, and to propose local adaptations. Additionally, communication between local platforms gives a global vision of the context permitting to have a global measure of the context and adapted reactions.

Each platform has three main tasks to accomplish:

-   *Capture of the context*: This task is important and implements tools to capture information of layer 1 (see Figure 3).
-   *Context Management Service.* Its role is to manage and evaluate information from layer 1 in order to evaluate if adaptation is required.
-   *Context Management Tools.* It proposes a set of mechanisms to adapt the application because of variations of the context.

The means operated to realize data flows #1 and #2 of Figure 9 depends on the types of context (Table 2). Interactions with local context use the mechanisms described in (Table 1) whereas those with distant context use services of the platform. The middleware of the platform offers services for context capture providing contextual information completed by time and localisation parameters as described in Figure 4.

| Flow | Type of context | | | Context |
|---|---|---|---|---|
| | **Hardware** | **User** | **Environment** | |
| #1 | System and network primitives | Interfaces | Sensors | Local |
| | Services of the platform | Services of the platform | Services of the platform | Distant |
| #2 | Resource allocation | Interfaces | Actuators | Local |
| | Services of the platform | Services of the platform | Services of the platform | Distant |

Table 2 : Interactions between Application and Context with a Platform

The role of the platform in this kind of organisation becomes central. We will now define more precisely the role and the architecture of a platform.

## 3.2  The platforms

Generally, we consider a platform as a set of elements of virtualization (Figure 10) allowing application designers to have a runtime environment independent of the hardware and network infrastructures, supporting distribution and offering non functional general services (persistence, security, transactions …) or services specific to a domain (business, medical …).

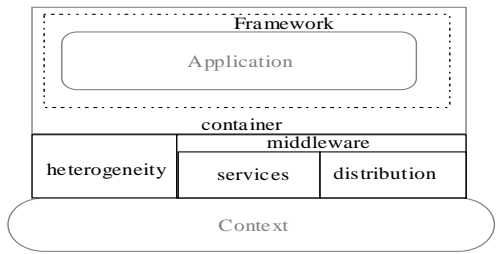

Figure 10 : Elements of virtualization in a platform

The container virtualizes the application or its components in make them suitable and compatible (interface) with the platform. The framework finishes this task allowing the designer to respect the corresponding model. The middleware virtualizes communications and offers services called by the application in order to access to the context. At last heterogeneity consists in virtualization of the hardware and the operating systems on witch the application runs.

Interactions between platform and application are bidirectional and represent the core aspect of the whole system (platform/application). The platform has its proper state evolving when modifications occur in the underlying level (context) and in the application. Consequently, the platform can trigger updates of the application state.

The interaction mode between application and platform can be achieved by:

-    service
-    container

In the first case, the changes of the state of the application that the platform knows are those inserted into the application itself by services, API or middleware calls ( *Figure 11* left), while in the second case the containers of the business components send to the platform information about their evolution ( *Figure 11* right). These containers can themselves offer some services to the business components or capture information about their changes of state by observing their behavior.

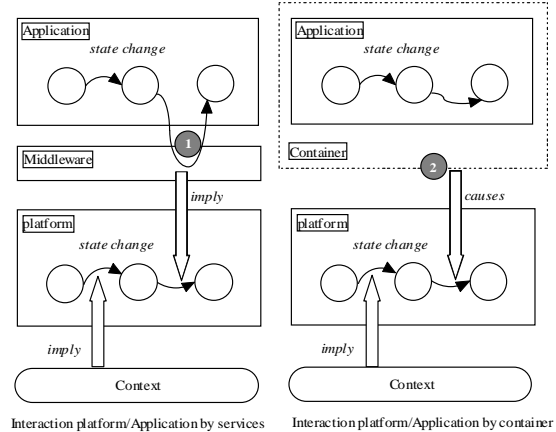

Interaction platform/Application by services    Interaction platform/Application by container

*Figure 11 : Modes of interaction between Application and Platform*

The interaction mode between platform and application allows distinguishing two families (Figure 12):

-    Non intrusive platforms;
-    Intrusive platforms.

A **non intrusive** platform acts on external elements of the application like data or uses a event based mechanism. It raises events when an internal state change occurs. These events can be caught by specific components of the application (event listeners). These modifications of external elements and these events imply the changes of the application state.

An **intrusive** platform can directly change the state of the application without participation of the application. This can be achieved by a direct action on the functional part either by a modification of the circulating of information either by directly modifying the architecture of the application itself. The use of objects and components facilitates greatly this task.

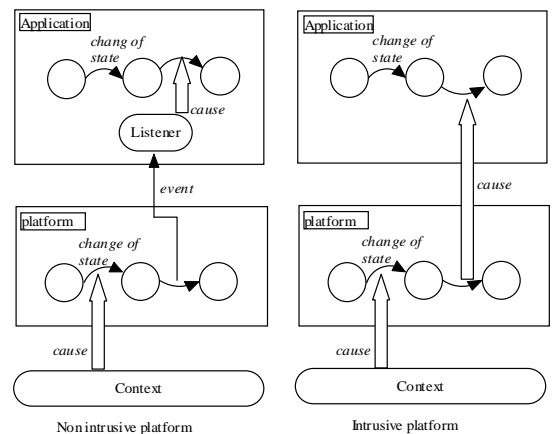

Non intrusive platform        Intrusive platform

Figure 12 : Modes of interaction between platform and application

## 3.3 Architecture of context aware adaptable applications

An overall schema of the architecture of an adaptable context aware application is presented in Figure 13. Relationship between platform and application are materialized by four flows:

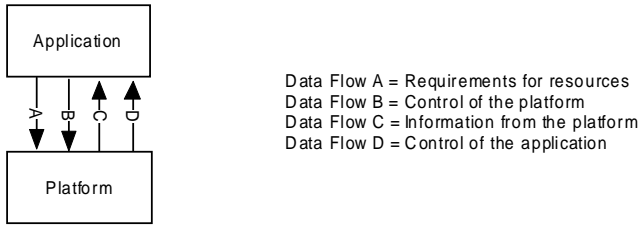

Data Flow A = Requirements for resources
Data Flow B = Control of the platform
Data Flow C = Information from the platform
Data Flow D = Control of the application

Figure 13: Information flows between application and platform [1]

This overall schema can be completed by adding the flows of interactions with the context as presented in Figure 9. We then obtain the general architecture shown in Figure 14 :

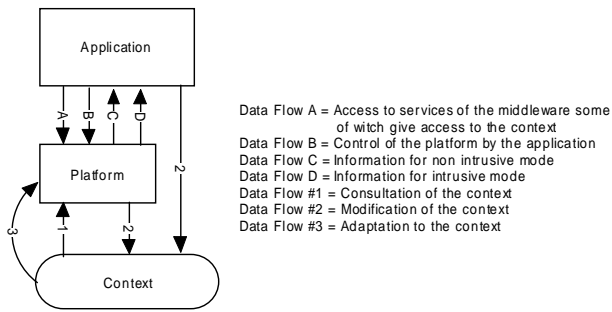

Data Flow A = Access to services of the middleware some
                 of wich give access to the context
Data Flow B = Control of the platform by the application
Data Flow C = Information for non intrusive mode
Data Flow D = Information for intrusive mode
Data Flow #1 = Consultation of the context
Data Flow #2 = Modification of the context
Data Flow #3 = Adaptation to the context

Figure 14 : Interactions between application, platform and context

Interactions between application and platform can be described as follow:

- Data Flow A corresponds to information from the application to the platform through usage of services of the middleware.
- Data Flow B represents the possibility to the application to configure the behavior of the platform (events priorities, filtering of contextual information, etc.)
- Data Flow C corresponds to the non intrusive mode of interaction between platform and application. It deals with events produced by the platform for the listeners inside the application.
- Data Flow D represents the intrusive mode of interaction between platform and application. It deals with updates of the application by the platform (modification of the architecture by adding/suppressing/moving components or by changing their business part).

Now, let's have a look on different context aware applications types that can be build according to data flows really used. Firstly it is important to notice that for context aware applications, data flow A is essential. In order to be adaptable, at least flow C or flow D need to be provided. If not, the platform is the only one able to be adaptable. The optional data flow B represents the possibility that the application has to configure the interaction modes corresponding to the flows A, C and D.

The Table 3 presents the four models of adaptation that it is possible to realize according to the flows used:

|   | Flows used | Type of interaction | Consequence |
|---|---|---|---|
| 1 | A | The platform is a middleware (services for accessing to local and distant context) | Only the platform is able to adapt itself to the context |
| 2 | A and C | The platform is a middleware (services for accessing to local and distant context) and offers an adaptation service | Adaptation is decided by the application according to information send by the platform |
| 3 | A and D | The platform is a middleware (services for accessing to local and distant context) and supervises the adaptation | Adaptation fully supervised |
| 4 | A, C and D | The platform is a middleware (services for accessing to local and distant context) and offers an adaptation service | Adaptation is partially supervised and partially decided by the application |

Table 3 : Possible models of adaptation according to the flows used

Data flow B allows to enrich the interaction types presented in the above Table 3:

- In the first case: the application only can configure the services of context access provided by the platform
- In the second case: the application can also choose the events which are indicated to it and their priority.
- In the third case: the application can configure the level of intrusion of the platform and eventually protect itself from it at some moments.
- The fourth case is the union of the two before.

According to the taxonomy proposed in [23] , middlewares like Aura [6] [7] [8] [9] , CARMEN [10] , CORTEX [11] [12] and CARISMA [13] [14] [15] belong to the first category while Cooltown [16] [17] , GAIA [18] [19] and MiddleWhere [20] belong to the second category. SOCAM [21] and Cadecomp [24] belong to the third category while MADAM [25] and Mobipads [22] belong to the fourth.

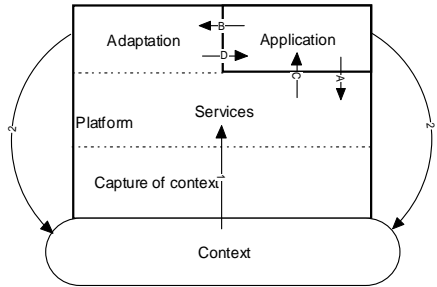

Figure 15: General schema of adaptation with a platform

We can then draw a general schema of an adaptable context aware application (Figure 15). The platform is distributed on every device hosting components of the applications. Then it can access to all contextual information. It offers a set of services in order to allow the application accessing to local or distant context (data flow A). Moreover it includes an adaptation manager sending events (data flow C) and a manager supervising the application (data flow D). The execution of this supervision manager can be configured by the application (data flow B).

### 3.4  Functional model of adaptation

The execution of an adaptable context aware application looks like a looped system: the context modifies the application, the execution of the application modify the context and so on. When a platform is introduced between the context and the application, a new loop appears because the platform itself is modified by the context and reciprocally, the platform modifies the context. Depending on using an intrusive or a non intrusive platform model, these loops are achieved by different data flows.

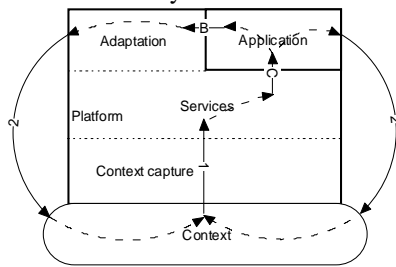

Figure 16 : Non intrusive adaptation model

−  Case 1: Adaptation controlled by the application (non intrusive model) :
   The context is captured by the platform (data flow #1) which signals its modifications to the application (data flow C). The application adapts itself using or not the services of the platform (data flow B). Activity of the application and platform modifies the context (data flow #2)

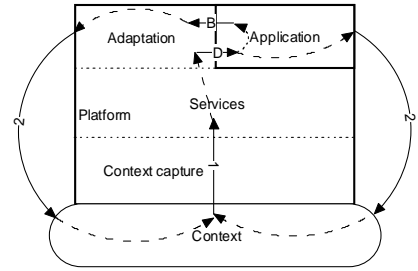

Figure 17 : Intrusive adaptation model

−  Case 2: Adaptation monitored by the platform (intrusive model):
−  The context is captured by the platform (data flow #1) which modifies the application (data flow D). This mechanism can be monitored by the application (data flow B). Activity of the application and of the platform modifies the context (data flow #2).

### 3.5  General architecture of a platform for adaptable context aware applications

The platform is composed of three main parts:
1. The capture of context is done by usual mechanisms as described in Table 1. They are system and network primitives, information system and sensors. Moreover, the platform also receives information about the application's running context from the containers of the business components (Figure 10).
2. The services concern both the application and the platform itself (more precisely the part in charge of the adaptation):
   For the application it corresponds to:
   • Services for accessing to the context (hardware, user, environment) with filtering possibilities (time, localisation)
   • Other usual services (persistence, …)
   For the adaptation it means:
   • Services for accessing to the context
   • Services for Quality of Service measurement
   • Services for reflexivity that is to say the knowledge that the system constituted by the platform and the application has of itself.
3. The adaptation matches the general schema of adaptation proposed in [3] which distinguishes two parts:
   • The evolution manager which implements the mechanisms of the adaptation;
   • The adaptation manager which monitors and evaluates the application.

IJCSI

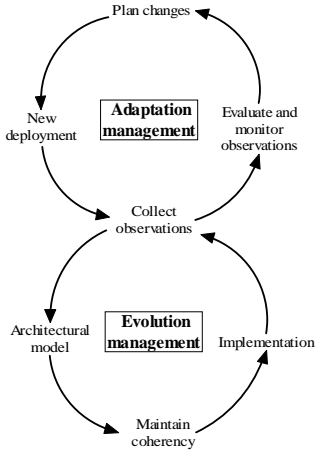

Figure 18 : General schema of adaptation [3]

The evolution manager monitors the application and its environment. Its architectural model selects an implementation maintaining the coherency of the application. The essential role of this manager is to check if deployment of the application is "causally connected" to the system [5]. Such a model integrates reflexivity like defined in [4] but limited to the architecture of the application and therefore protecting the encapsulation of the business components. The adaptation manager receives observations measured by the evolution manager. It evaluates them in order to select an adaptation and to find a new deployment of the components of the application (Figure 18).

## 4. Kalimucho platform and implementation tools

The architecture of the application has to be virtualized in order to be monitored by the platform. The general architecture of the *Kalimucho* platform is the following:

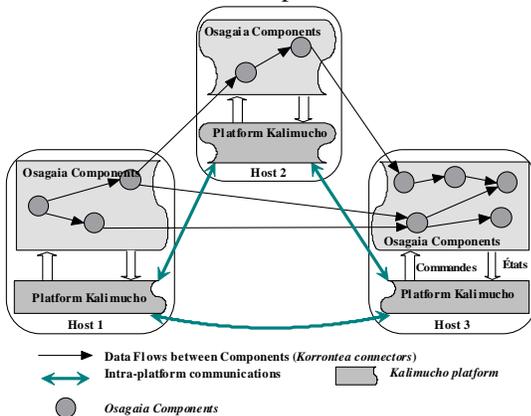

Figure 19: Kalimucho's General Architecture

It is based on a distributed service based platform implementing non-functional services for adaptations (*layer 2* – Figure 3). The functional part is implemented with software and hardware components running into the generic *Osagaia* container. Communication between components uses the generic framework called *Korrontea*. This framework is a first class component connector able to implement various communications policies.

### 4.1 Kalimucho architecture

We propose to build the architecture of adaptable context aware applications on a distributed platform called *Kalimucho*.

The application is made of business components (BC) interconnected by information flows. To directly modify the architecture of the application it is necessary that the platform should be able to add/remove/move/connect/disconnect the components. Moreover the platform has to capture the context on every site. We created a container for information data flows named *Korrontea* and another for business components named *Osagaia* [26] . These containers collect local contextual information from business components and connectors and send them to the platform. They receive back supervisions commands from the platform. Interactions between the platform and the application are implemented with the flows shown in Figure 20. We can notice that because *Korrontea* containers have a non functional role into the application (information transportation), they do not accept the data flow C and are not event listeners. On the other hand, some BC can react to context events sent by the platform towards *Osagaia* containers.

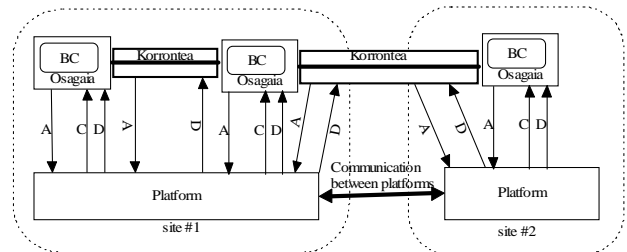

Figure 20 : Interactions between application and platform in Kalimucho

Our work deals with various devices as sensors (which are CLDC compliant), PDA, SmartPhones (CDC compliant) and traditional PCs. Such an heterogeneous environment implies several services variations devoted to the platform: The capture of the context is done by services (*Osagaia*) and flow containers (*Korrontea*). Depending on the host running the component, it will capture users, environment, hardware, temporal or geographic information (see layer 1 - Figure 3). The second layer (*context management services*) is done by implementing an heuristic in order to evaluate the current Quality of Service (QoS) and to propose adaptations if needed and if possible. The last layer (*context management tools*) gives solutions to provide adaptations (add/remove/move/connect/disconnect components).

The platform is distributed on every machine on which components of the application are deployed (desktops, mobile devices and sensors). The different parts of the platform communicate through the network. Communications between BCs (local or distant) are achieved by data flows encapsulated into *Korrontea* containers.

Various versions of the platform are implemented on the different hosts according to their physical capacities. On a desktop all the parts of the platform are implemented whereas, on a mobile device, and particularly on a wireless sensor, light versions are proposed (one for CDC and one for CLDC compliant hosts). Consequently, only non avoidable services for the host are deployed (for example a service for persistence is useless on a sensor). In the same way, the adaptation manager implemented on a mobile device can be lightened using internal services of one of the neighbouring platform (for example, only local routing information is available on a limited device). If the platform of this device needs to find others routes in order to set up a new connection, it has to use services of the platforms implemented on neighbouring desktops.

## 4.2 Osagaia Software Component Model

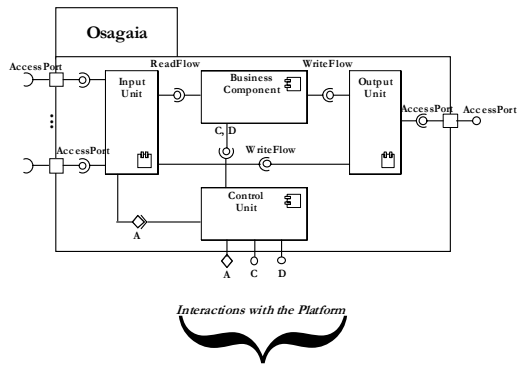

Figure 21: *Osagaia* Conceptual Model

Finally we design the software component model in order to ensure the implementation of distributed applications according to the specifications expressed by functional graphs [41] .

Functional components are called business component since they implement the business functionalities of applications. These components need to be executed into a container whose role is to provide non-functional implementation for components. The architecture of this container is shown in Figure 21, we call it *Osagaia*. Its role is to perform interactions between business components and their environment. It is divided into two main parts: the exchange unit (composed of input and output units, see Figure 21) and the control unit. The exchange unit manages data flows input/output connections. The control unit manages the life cycle of the business component and the interactions with the runtime platform. Thus, the platform supervises the containers and, indirectly, the business components (a full description of the Osagaia software component model is available in [31] ). Thanks to this container, business components read and write data flows managed by Connectors called Korrontea (see Figure 22). Its main role is to connect software components of the applications. The Korrontea container receives data flows produced by components and transports them. It is made up of two parts. The control unit implements interactions between the Korrontea container and the platform while an exchange unit manages the input/output connections with components. The container is the distributed entity of our model, i.e. it can transfer data flows between different sites of distributed applications. The flow management is done according to the business part of the connector implementing both the communication mode (client/server for example) and the communication politic (with or without synchronization, loss of data, etc.). A full description of the Korrontea component model is available in [28] ).

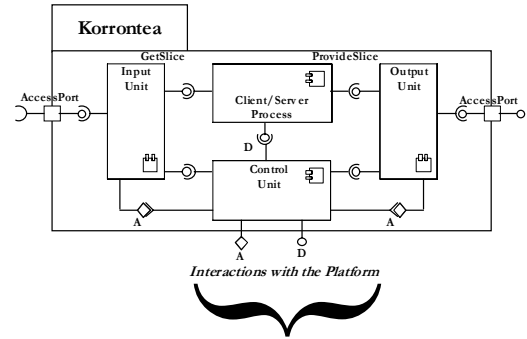

Figure 22: *Korrontea* Conceptual Model

## 5. Conclusion

In this paper, we presented an overview of adaptable applications. Because such applications need knowledge of their environment, we made a definition of the context and presented it according to applications uses. Next, we present adaptation management politics and their possible implementation. This part was followed by a presentation of implementation tools able to provide adaptations. We finished by the description of the *Kalimucho* platform, software and connectors containers models used in order to make adaptations.

Implementing context-aware adaptable applications with a platform helps having a global view of the application and of the context. The global view of the application permits an optimum mobility and resource management. The global view of context permits considering the whole context of the application instead of the only local one.

The system composed of the platform and the application make up a reflexive context aware system.

The problem of such an approach is its inherent complexity. Context aware platforms become more and more complex in order to manage a context more and more variable and evanescent. So, depending on the targeted application, it could be much more interesting to provide various lighter, specialized and reflexive platforms providing a view of their state. Moreover, such platforms are able to be heaped with other light, specialized and reflexive ones.

The influence of the environment on the system behavior leads to strongly couple the execution platform and the application[38] . So design methods for applications and platforms must also be coupled to constitute a sole design method.

Instead of making a whole design step, we propose a life-cycle including both application and platform (which is also an application – this is **recursive**) to finish with implementation tools (platform specific, component and connector models and specific implementations). Such approach let us imagine wide development with automatic code generation.

# 6.  Bibliography

[1]    Weiser, M. (1991) 'The computer for the 21st century', Scientific American, pp.94–104.

[2]    C. Efstratiou, K. Cheverst, N. Davies, A. Friday : "An Architecture for the Effective Support of Adaptive Context-Aware Applications". In Proc. of the Second Int'l Conference on Mobile Data Management (MDM 2001).

[3]    P. Oreizy, M. M. Gorlick, R. N. Taylor, D. Heimbigner, G. Johnson, N. Medvidivic, A. Quilici, D. S. Rosenblum, A. L. Wolf : "An architecture-based approach to self-adaptative software". IEEE Intelligent Systems, vol 14 n°3, pp : 54-62. Mai/Juin 1999.

[4]    P. Maes : "Concepts and experiments in computational reflection". In proceedings of the conference on object-oriented systems, languages and applications (OOPSLA'87), pp : 147-155. Orlando, Florida 1987.

[5]    S. Krakowiak : "Introduction à l'intergiciel", Intergiciel et construction d'application réparties (ICAR), pp : 1-21, 19 Janv 2007, Licence Creative Commons

[6]    D. Garlan, D. Siewiorek, A. Smailagic, and P. Steenkiste. Project Aura: Toward Distraction-Free Pervasive Computing. IEEE Pervasive computing, 1(2):22–31, April–June 2002.

[7]    U. Hengartner and P. Steenkiste. Protecting access to people location information. In D. Hutter, G. Müller, W. Stephan, and M. Ullmann, editors, SPC, volume 2802 of LNCS, pages 25–38. Springer, 2003.

[8]    G. Judd and P. Steenkiste. Providing contextual information to pervasive computing applications. In PERCOM '03: Proceedings of the First IEEE International Conference on Pervasive Computing and Communications, page 133,Washington, DC, USA, 2003. IEEE Computer Society.

[9]    J. P. Sousa and D. Garlan. Aura: An architectural framework for user mobility in ubiquitous computing environments. In WICSA 3: Proceedings of the IFIP 17th World Computer Congress - TC2 Stream / 3rd IEEE/IFIP Conference on Software Architecture, pages 29–43, Deventer, The Netherlands, The Netherlands, 2002. Kluwer, B.V.

[10]   P. Bellavista, A. Corradi, R. Montanari, and C. Stefanelli. Context-aware middleware for resource management in the wireless internet. IEEE Transactions on Software Engineering, 29(12):1086–1099, 2003.

[11]   H. A. Duran-Limon, G. S. Blair, A. Friday, P. Grace, G. Samartzisdis, T. Sivahraran, and M. WU. Contextaware middleware for pervasive and ad hoc environments, 2000.

[12]   C.-F. Sørensen, M. Wu, T. Sivaharan, G. S. Blair, P. Okanda, A. Friday, and H. Duran-Limon. A context-aware middleware for applications in mobile ad hoc environments. In MPAC '04: Proc. of the 2nd workshop on Middleware for pervasive and ad-hoc computing, pages 107–110, New York, NY, USA, 2004. ACM Press.

[13]   L. Capra. Mobile computing middleware for context aware applications. In ICSE '02: Proceedings of the 24th International Conference on Software Engineering, pages 723–724, New York, NY, USA, 2002. ACM Press.

[14]   L. Capra, W. Emmerich, and C. Mascolo. Reflective middleware solutions for context-aware applications. Lecture Notes in Computer Science, 2192:126–133, 2001.

[15]   L. Capra, W. Emmerich, and C. Mascolo. Carisma: context-aware reflective middleware system for mobile applications. IEEE Transactions on Software Engineering, 29(10):929 – 45, 2003/10/.

[16]   J. Barton and T. Kindberg. The Cooltown user experience. Technical report, Hewlett Packard, February 2001.

[17]   P. Debaty, P. Goddi, and A. Vorbau. Integrating the physical world with the web to enable context-enhanced services. Technical report, Hewlett-Packard, Sept. 2003.

[18]   M. Roman, C. Hess, R. Cerqueira, A. Ranganathan, R. Campbell, and K. Nahrstedt. A middleware infrastructure for active spaces. IEEE Pervasive Computing, 1(4):74 – 83, 2002/10/.

[19]   M. Román, C. K. Hess, R. Cerqueira, A. Ranganathan, R. H. Campbell, and K. Nahrstedt. Gaia: A Middleware Infrastructure to Enable Active Spaces. IEEE Pervasive Computing, pages 74–83, Oct–Dec 2002.

[20]   A. Ranganathan, J. Al-Muhtadi, S. Chetan, R. H. Campbell, and M. D. Mickunas. Middlewhere: A middleware for location awareness in ubiquitous computing applications. In H.-A. Jacobsen, editor, Middleware, volume 3231 of Lecture Notes in Computer Science, pages 397–416. Springer, 2004.

[21]   T. Gu, H. K. Pung, and D. Q. Zhang. A middleware for building context-aware mobile services. In Proceedings of IEEE Vehicular Technology Conference, May 2004.

[22]   A. Chan and S.-N. Chuang. Mobipads: a reflective middleware for context-aware mobile computing. IEEE Transactions on Software Engineering, 29(12):1072 – 85, 2003/12.

[23]   Kristian Ellebæk Kjær. A survey of context-aware middleware. In Proceedings of the 25th conference on

IASTED International Multi-Conference: Software Engineering Innsbruck, Austria ,Pages 148-155, 2007

[24] Dhouha Ayed, Nabiha Belhanafi, Chantal Taconet, Guy Bernard. Deployment of Component-based Applications on Top of a Context-aware Middleware. - The IASTED International Conference on Software Engineering (SE 2005) - Innsbruck, Austria - February 15-17, 2005. http://picolibre.int-evry.fr/projects/cadecomp

[25] MADAM Consortium. MADAM middleware platform core and middleware services. Editor Alessandro Mamelli (Hewlett-Packard), deliverable D4.2, 30 March 2007. hppt://www.intermedia.uio.no/confluence/madam/Home

[26] C. Louberry, M. Dalmau, P. Roose – Architectures Logicielles pour des Applications Hétérogènes Distribuées et Reconfigurables – NOTERE'08 - 23-27/06/2008, Lyon.

[27] Robert Laddaga, Paul Robertson, Self Adaptive Software: A Position Paper, SELF-STAR: International Workshop on Self-* Properties in Complex Information Systems, 31 May - 2 June 2004

[28] Holger Schmidt, Franz J. Hauck: SAMProc: Middleware for Self-adaptive Mobile Processes in Heterogeneous Ubiquitous Environments. 4th Middleware Doctoral Symposium - MDS, co-located at the ACM/IFIP/USENIX 8th International Middleware Conference (Newport Beach, CA, USA, November 26, 2007).

[29] Baresi, L.; Baumgarten, M.; Mulvenna, M.; Nugent, C.; Curran, K.; Deussen, P.H. - Towards Pervasive Supervision for Autonomic Systems - Distributed Intelligent Systems: Collective Intelligence and Its Applications, 2006. DIS 2006. IEEE Workshop on Volume, Issue, 15-16 June 2006 Page(s):365 – 370.

[30] Emmanuel bouix, Philippe Roose, Marc Dalmau - The Korrontea Data Modeling - Ambi Sys 2008 - International Conference on Ambient Media and Systems - 11/14 february, Quebec City, Canada, 2008.

[31] E. Bouix, M. Dalmau, P. Roose, F. Luthon. A Component - Model for transmission and processing of Synchronized Multimedia Data Flows. In Proceedings of the 1st IEEE International Conference on Distributed Frameworks for Multimedia Applications (France, February 6-9 2005).

[32] D. Garlan, J. Kramer, and A. Wolf, editors. Proceedings of the First ACM SIGSOFT Workshop on Self-Healing Systems (WOSS '02). ACM Press, 2002.

[33] [Roman, 2000] Roman G.C., Picco, G.P. Murphy A.L. – Software Engineering for mobility : a roadmap – ICSE 2000 – ACM Press, New York, USA, p. 241-258 – 2000.

[34] A.K. Dey G.D. Abowd – Towards a better understanding of context and context-awareness – CHI 2000 - Workshop on the What, Who, Where, When and How of Context-Awareness, The Hague, Netherlands, April 2000.

[35] Dey, A.K. and Abowd, G.D. 'A conceptual framework and a toolkit for supporting rapid prototyping of context-aware applications', HCI Journal, Vol. 16, Nos. 2–4, pp.7–166.

[36] T. Chaari, F. Laforest - L'adaptation dans les systèmes d'information sensibles au contexte d'utilisation: approche et modèles. Conférence Génie Electrique et Informatique (GEI), Sousse, Tunisie, mars 2005. pp. 56-61.

[37] Matthias Baldauf, Schahram Dustdar, Florian Rosemberg - A survey on context-aware systems – Int'l journal on Ad Hoc and Ubiquitous Computing, Vol.2, N°4, 2007.

[38] T.A. Henzinger and J. Sifakis. The Embedded Systems Design Challenge *Invited Paper, FM 2006, pp. 1-15*.

[39] Indulska, J. and Sutton, P. (2003) 'Location management in pervasive systems', CRPITS'03: Proceedings of the Australasian Information Security Workshop, pp.143–151.

[40] A. Ranganathan, J. Al-Muhtadi, S. Chetan, R. H. Campbell, and M. D. Mickunas. Middlewhere: A middleware for location awareness in ubiquitous computing applications. Vol. 3231 of LNCS, pages 397–416. Springer, 2004.

[41] Sophie Laplace, Marc Dalmau, Philippe Roose - Kalinahia: Considering quality of service to design and execute distributed multimedia applications - NOMS 2008 - IEEE/IFIP Int'l Conference on Network Management and Management Symposium - 7-11/04/2008 Salvador de Bahia, Brazil, 2008.

[42] Bill Schilit, Marvin Theimer - Disseminating Active Map Information to Mobile Hosts - IEEE Network, September, 1994

[43] Jason Pascoe , Nick Ryan, David - Using while moving: HCI issues in fieldwork environments –ACM Transactions on Computer-Human Interaction (TOCHI) Vol. 7 , Issue 3 (09/2000) - Special issue on human-computer interaction with mobile systems - 2000

[44] K.E. Kjær - A Survey of Context-Aware Middleware - Software Engineering - SE 2007 - Innsbruck, Austria, 2007.

[45] Frédérique Laforest - De l'adaptation à la prise en compte du contexte – Une contribution aux systèmes d'information pervasifs – Habilitation à Diriger les Recherches, Université Claude Barnard, Lyon I, 2008.

[46] Daniel Cheung-Foo-Wo, Jean-Yves Tigli, Stéphane Lavirotte, Michel Riveill. "Contextual Adaptation for Ubiquitous Computing Systems using Components and Aspect of Assembly" in Proc. of the Applied Computing (IADIS), IADIS, Salamanca, Spain, 18-20 feb 2007

[47] Guanling Chen, David Kotz - A Survey of Context-Aware Mobile Computing Research - Dartmouth College Technical Report TR2000-381, November 2000.

[48] H. Lieberman and T. Selker - Out of context: Computer systems that adapt to, and learn from, context – IBM System Journal - Volume 39, Numbers 3 & 4, MIT Media Laboratory 2000.

[49] Pierre-Charles David, Thomas Ledoux - WildCAT: a generic framework for context-aware applications, Proceedings of the 3rd international workshop on Middleware for pervasive and ad-hoc computing, ACM International Conference Proceeding Series; Vol. 115

[50] A. Harter, A. Hopper – A distributed location system for the active office. IEEE Networks, 8(1):6270, 1994.

[51] Want, R. Schilit, B.N. Adams, N.I. Gold, R. Petersen, K. Goldberg, D. Ellis, J.R. Weiser, M. - An overview of the PARCTab ubiquitous computing environment. IEEE Personal Communications, 2(6): 2833, 1995.

**Marc Dalmau** is an IEEE member and Assistant Professor in the department of Computer Science at the University of Pau, France. He is a member of the TCAP project. His research interests include wireless sensors, software architectures for distributed multimedia applications, software components, quality of service, dynamic reconfiguration, distributed software platform, information system for multimedia applications.

**Philippe Roose** is an Assistant Professor in the department of Computer Science at the University of Pau, France. He is responsible of the *TCAP project - Video flows transportation on sensor networks for on demand supervision.* His research interests include wireless sensors, software architectures for distributed multimedia applications, software components, quality of service, dynamic reconfiguration, COTS, distributed software platform, information system for multimedia applications.

**Sophie Laplace** is Doctor Biographies in the department of Computer Science at the University of Pau, France. Her researches interests include formal methodology, Quality of Service design and evaluation. Her works mainly focus on multimedia applications. She defended her PhD (*Software Architecture Design in order to integrate QoS in Distributed Multimedia Applications*) thesis in 2006